\documentclass[pra, onecolumn, nofootinbib]{revtex4}
\usepackage[dvips]{graphics}
\usepackage{amssymb}
\usepackage{epsfig}
\usepackage{subfigure}

\begin{document}
\date{\today}
\title{Relaxation Phenomena in a System of Two Harmonic Oscillators}
\author{Antonia Chimonidou}
\email[email:]{antonia@physics.utexas.edu} \affiliation{The University of Texas at Austin, Center for Complex
Quantum Systems, 1 University Station C1600, Austin TX 78712}
\author{E.~C.~G. Sudarshan}
\affiliation{The University of Texas at Austin, Center for Complex Quantum Systems, 1 University Station C1600,
Austin TX 78712}
\begin{abstract}
We study the process by which quantum correlations are created when an interaction Hamiltonian is repeatedly
applied to a system of two harmonic oscillators for some characteristic time interval. We show that, for the
case where the oscillator frequencies are equal, the initial Maxwell-Boltzmann distributions of the uncoupled
parts evolve to a new Maxwell-Boltzmann distribution through a series of transient Maxwell-Boltzmann
distributions, or quasi-stationary, non-equilibrium states. Further, we discuss why the equilibrium reached when
the two oscillator frequencies are unequal, is not a thermal one. All the calculations are exact and the results
are obtained through an iterative process, without using perturbation theory.
\end{abstract}

\maketitle
\section{Introduction}
The interaction between two isolated systems can be a resource for quantum information processing
\cite{Nielsen}. However, the interaction between an isolated system and the environment surrounding it can lead
to decoherence, an undesirable loss of information that was initially available. One way or the other, any
interaction between two initially uncoupled subsystems leads to the exchange of quantities such as purity or
polarization \cite{Cesar}, or for thermodynamical systems, temperature. To understand and to control this
interaction process, as well as to prevent decoherence, we need to study the mechanism by which this exchange
occurs. There has been a lot of interest and work in this problem \cite{Munro, Vidal, Vedral, Wootters, Wei,
Raimond}. Most of this work involves the coupling between two (or more) two-level quantum systems, or qubits,
the fundamental units used in quantum computing \cite{Zurek, Wootters2, Saguia, Rossini}. In this paper, we will
instead concentrate on the interaction between two harmonic oscillators. More specifically, we are interested in
calculating the time evolution of each oscillator separately, after an external interaction Hamiltonian has been
repeatedly applied to the bipartite system for enough successive characteristic time intervals, until a new
equilibrium has been reached.

Consider a bipartite system $\rho_{12}(0)$ composed of two initially uncoupled subsystems $\rho_1(0)$ and
$\rho_2(0)$. If an interaction Hamiltonian is applied to this system for a characteristic time interval $\tau$,
the two subsystems will interact and evolve to two new states $\rho_1(\tau)$ and $\rho_2(\tau)$. In this paper,
we will follow a program in which our system is ``refreshed'' after each characteristic time interval $\tau$,
and the same interaction Hamiltonian is repeatedly applied to it until the new equilibrium is reached. The
``refresh'' procedure is a very important ingredient in our model and we will spend some time explaining it in
detail in Secs. II and III.

This article is structured as follows: in Sec. II, we develop a general program for the relaxation of a
bipartite system by introducing a periodically repeated interaction scheme. In Sec. III, we apply this program
to a system of two initially uncoupled quantum harmonic oscillators. In Sec. IV, we discuss the physical
interpretation and consequences which arise from solving this problem and present the conditions needed for a
new equilibrium to be attained. We explore specific cases in some detail and present our results graphically.
Finally, in Sec. V, we discuss the implications of our results and suggest possible future directions to be
explored.

\section{General Program for the Evolution of a Bipartite System}
To quantitatively describe physical situations with pure as well as mixed ensembles, we use the density matrix
formalism, introduced independently by Landau and von Neumann in the $1920$'s \cite{Neumann, Landau}. Density
matrices not only portray the probabilistic nature of a quantum system, but they also contain all the physically
significant information we can possibly obtain about the ensemble in question. The time evolution of a closed
quantum system $\rho(t)$ governed by a time-independent Hamiltonian $\hat{H}$ is represented by:
\[
\rho(t)=\hat{U}(t,t_0)\rho(t_0)\hat{U}^\dag(t,t_0),
\]
where $\hat{U}(t,t_0)$ is the unitary time evolution operator given by $\hat{U}(t,t_0)=e^{-i\hat{H}t/\hbar}$.
Quantum correlations are generated when two such systems are forced to interact through some interaction
Hamiltonian $\hat{H}_{int}$.

Consider a bipartite system $\rho_{12}(0)$ composed of two initially uncoupled subsystems $\rho_1(0)$ and
$\rho_2(0)$. Although the evolution of $\rho_{12}(0)$ is unitary, the reduced evolution of $\rho_1(0)$ or
$\rho_2(0)$ is in general not unitary \cite{notunit}. Knowing the form of $\rho_1$ at any time $t_1$ is not
sufficient to predict its form at a future time $t_2$. So to extract the reduced evolution of subsystem $1$, we
need to first calculate the time evolution of $\rho_{12}(0)$ and eliminate the effect of the unwanted part by
taking the partial trace of the evolved complete system over subsystem $2$. Namely:
\[
\rho_1(t)=\mbox{Tr}_{2}[\hat{U}(t)\rho_{12}(0)\hat{U}^\dag(t)],
\]
where $\rho_{12}(0)=\rho_1(0)\otimes\rho_2(0)$ and $\hat{U}(t)=e^{-i\hat{H}t/\hbar}$, with $\hat{H}$ being the
total Hamiltonian of the composed system, including $\hat{H}_{int}$.

In our work, we will follow a scheme similar to one proposed by Rau \cite{Rau}, in which a constant interaction
Hamiltonian $\hat{H}_{int}$ is applied to a bipartite system $\rho_{12}(0)=\rho_1(0)\otimes\rho_2(0)$ composed
of two initially uncoupled harmonic oscillators with density matrices $\rho_1(0)$ and $\rho_2(0)$, and with
$\rho_2$ kept in an inexhaustible temperature bath. In Rau's model, emphasis is given to studying how system $1$
relaxes when it is in contact with system $2$. The interaction Hamiltonian is applied to the composite system
$\rho_{12}(0)$ for a constant characteristic time interval $\tau$, interrupted, and then reapplied for another
time interval $\tau$ to the time-evolved state $\rho_{12}(\tau)=\rho_1(\tau)\otimes\rho_2(\tau)$, where
$\rho_1(\tau)=\mbox{Tr}_2[\hat{U}(\tau)\rho_{12}(0)\hat{U}^\dag(\tau)]$, and $\rho_2(\tau)=\rho_2(0)$. As
mentioned above, system $2$ is assumed to be in contact with an inexhaustible temperature bath, or reservoir, so
that any change it may undergo during the time interval $\tau$ is negligible. In other words, at $t>0$, a
general, constant interaction Hamiltonian is applied for a characteristic time interval $\tau$, the system is
allowed to interact, it is then``refreshed'', and the process is repeated. The ``interaction'' process is
defined as follows: the two quantum systems $\rho_1(0)$ and $\rho_2(0)$ interact with one another through the
interaction Hamiltonian $\hat{H}_{int}$ during a time interval $0<t<\tau$, where $\tau$ is fixed. In Rau's
model, the ``refresh'' process is the extraction of the reduced time-evolved quantum part $\rho_1(\tau)$ from
the complete system $\rho_{12}(\tau)$ and the resetting of $\rho_2(\tau)$ back to its initial form. Finally, the
``repeat'' process is the application of the same interaction Hamiltonian to the new state
$\rho_{12}(\tau)=\rho_1(\tau)\otimes\rho_2(\tau)$ (with $\rho_2(\tau)=\rho_2(0)$) for another time interval
$\tau<t<2\tau$. There are several areas in physics where such a model is a good approximation. For example, in
spintronics, in the case of paramagnetic relaxation, a system of spins interacts with a lattice, and the
information is encoded in the spin states of the conduction electrons \cite{Spintronics}. In this case, the
conduction electron takes the role of system $1$, and the atoms in the lattice take the role of system $2$, or
the environment. As the conduction electron passes through the lattice, it interacts with a lattice atom, and as
a consequence, its state is changed. This new state will now interact with another atom in the lattice,
identical to the first one.

The following assumptions are made in the formulation of generating relaxation as described in Rau's model:
First, it is assumed that there is statistical independence between the two density matrices of subsystems $1$
and $2$ at $t=0$, or in other words, the density matrix of the combined system can be factorized into density
matrices of the component subsystems at $t=0$, i.e., $\rho_{12}(0)=\rho_1(0)\otimes\rho_2(0)$. Second, the
relaxation time of the system of interest, subsystem $1$, is much larger than the characteristic interval
$\tau$. In other words, the system is reset before the time interval specified by the inverse of the interaction
energy. Third, time averages are never taken, but after every characteristic time interval $\tau$, partial trace
with respect to the reservoir is taken. Finally, system $2$ is assumed to be in an inexhaustible temperature
bath so that any change in it during the interaction process is negligible. At each time interval $\tau$, the
system goes back to its original state at $t=0$ while system $1$, the system of interest, changes its state. The
mechanism of refreshing introduced in this model annuls the correlation between the two systems $1$ and $2$ at
every characteristic time interval $\tau$ and replaces the time-evolved state of subsystem $2$ with an identical
copy of its original state.

Implicit in this model is the assumption, invoked by Pauli \cite{Pauli}, that the occupation numbers, which
correspond to the diagonal elements of the density matrix, remain good quantum numbers at all times. In other
words, it is assumed that, during the course of time, off-diagonal terms become unimportant. Pauli eliminated
them by invoking a ``random- phase approximation'' at all times. This approximation is a serious ingredient in
the analysis of this problem and we believe it is worthwhile to analyze its use at this point. To begin, we
stress that the disappearance of interferences between a system and its environment is an irreversible process.
Irreversible, or energy-dissipating processes always involve transitions between quantum states. Such processes
are described, at the simplest level, by master or rate equations. The Pauli master equation \cite{Pauli}, is
the most commonly used model of irreversible processes in simple quantum systems. It can be derived from
elementary quantum mechanics and by neglecting off-diagonal terms. Among other (inessential to our discussion)
approximations, the derivation of the Pauli master equation suffers from the restriction that one has to discard
any built-up of phase relations by invoking a repeated ``random-phase approximation'' at a series of times at
microscopically small intervals. This restriction may find its justification in Van Kampen's analysis of the
problem in deriving the master equation from quantum mechanics \cite{Van Kampen, Van Kampen2}. Specifically, Van
Kampen observed that, by writing a master equation, we only intend to get statements on macroscopically
observable properties of (statistically) large systems. In trying to derive a master equation from quantum
mechanics, we must therefore first construct, following Pauli, a suitable coarse-graining of phase-space in such
a way that the quantities of the statistical theory are only those that can be measured macroscopically. Van
Kampen's derivation of the master equation highlights quite explicitly the inherent difficulties of
non-equilibrium statistical mechanics. To derive a master equation, we must postulate a number of (justifiable)
mathematical assumptions, unfortunately in most cases without being able to give the explicit criteria on the
microscopic dynamics of the system for these assumptions to be valid. The general attitude is that because many
large systems evolve smoothly on a macroscopic time scale, microscopic details are most likely not important,
and must therefore be suppressed. Further, it is assumed that the evolution of the reduced density matrix is a
Markoff process. This translates to invoking a repeated random-phase approximation, i.e. neglecting or
suppressing any dynamical built-up of phases as time evolves. For example, in the case of photon scattering,
interference terms connecting different positions become unobservable at the macroscopic body itself, though
still existing in the whole system. This was first discussed by von Neumann in his theory of the measurement
process \cite{Neumann2}. Another possible physical mechanism which may aid in explaining the validity of Pauli's
approximation is the interaction of the system with its natural environment. It has been shown
\cite{Kubler,Zurek2}, that the interaction between the system and its environment causes some interference terms
to become unobservable.

In this paper, we will follow a scheme that deviates slightly from the one proposed by Rau. In our model, we
again start with two initially uncoupled subsystems described by density matrices $\rho_1(0)$ and $\rho_2(0)$.
The crucial difference between the program analyzed by Rau and what we propose here, is that system $2$ is no
longer kept in an inexhaustible temperature bath. After a constant interaction Hamiltonian is applied to the
bipartite system for a characteristic time interval $\tau$, the system is ``refreshed'', but the ``refresh''
procedure is no longer the same as that described above. Instead of concentrating our attention to studying how
system $1$ relaxes when in contact with system $2$ like in Rau's model, we will now look at how each system
relaxes when in contact with the other, once $\hat{H}_{int}$ is applied to the combined system. System $2$ will
no longer act as the reservoir which goes back to its initial state after every $\tau$. Instead, both subsystems
will be equally important and both will undergo a change in their states during the interaction process. We will
no longer be concentrating on a system in contact with an unchanging environment, but in two systems of interest
in contact with each other.

The ``refresh'' procedure as defined in our model can be written as a transformation $\rho_{12}\rightarrow
Tr_2[\rho_{12}]\otimes Tr_1[\rho_{12}]$. As an aside that will become transparent later on, we state at this
point, that even though the ``refresh'' procedure is expressed as the tensor product between the time-evolved
reduced subsystem states obtained by performing partial traces over each unwanted part separately, the operation
of taking the partial trace is performed only once, and over only one of the oscillator subsystems. The partial
trace operation assumes that we consider the relative phase between systems $1$ and $2$ to be completely
randomized and thus we can integrate over it. In addition to this, we assume that we know nothing about the
occupation numbers of each mode of the second system and hence we can sum over them when computing the partial
density matrix of system $1$ (and vice-versa for system $2$). Further, we assume statistical independence
between the two density matrices $\rho_1$ and $\rho_2$ at each time $\tau$. For the time being, we simply state
that the ``refresh'' procedure is composed of nothing but a single spectrum measurement in the number basis of
one of the oscillators. This single measurement has a twofold effect: it diagonalizes the output global state in
\emph{both} oscillator number bases $\{|n_1\rangle\}$ and $\{|n_2\rangle\}$, and at the same time decouples the
two oscillators by allowing us to write the global output state as a tensor product of the two reduced,
post-measurement states. The physical implementation of our model along with this last assumption will become
clear in Sec. III, where we explicitly calculate the reduced density matrix $\rho_1(\tau)$, and extensively
discuss the details of the ``refresh'' procedure.

Here, we must emphasize that the ``refresh'' procedure defined in our model is different from the Pauli
``random-phase approximation'' employed in Rau's model. Formally, the ``refresh'' procedure is a transformation
$\rho_{12}\rightarrow\mbox{Tr}_{2}[\rho_{12}]\otimes\mbox{Tr}_{1}[\rho_{12}]$ and appears to be nonlinear with
respect to $\rho_{12}$ (this can be easily observed if $\rho_{12}$ is written as a convex sum of pure density
matrices). In contrast, the Pauli ``random-phase approximation'' is a transformation $\rho_{12}\rightarrow
\sum_{n_1n_2}|n_1n_2\rangle \langle n_1n_2|\rho_{12}|n_1n_2\rangle \langle n_1n_2|$ (with $|n_1n_2\rangle$ being
the orthonormal basis of the total free Hamiltonian), that is linear with respect to $\rho_{12}$, and
corresponds to vanishing of off-diagonal terms in $\rho_{12}$. Whereas the partial trace assumes that we have no
knowledge of the occupation numbers of each mode in the second subsystem and hence we end up averaging over
those occupation numbers, the ``random-phase approximation'' considers \emph{all} phase relationships to be
random. In contrast to Rau's model, we do not have a Markovian master equation for either of the two
oscillators. Since, in our case, the second system is not reset back to its initial state after each time
interval $\tau$, the other system keeps a memory of what happened in the previous step. The nonlinearity of the
``refresh'' procedure may raise philosophical questions on the physical realizations associated with it. At this
point, it is perhaps appropriate to clear any confusions that may arise due to the apparent nonlinearity of the
transformation: the ``refresh'' procedure is expressed as the tensor product between the two reduced
post-measurement states, written as states which are obtained by taking partial traces over the corresponding
unwanted parts. However, like mentioned above and discussed further in section III of this article, the
transformation can be obtained by making a \emph{single}, spectrum measurement on the number basis of just one
of the two oscillators. This measurement will not only determine the probability distribution of the occupation
numbers in oscillator $1$, but in addition, that of oscillator $2$. We strongly emphasize that the operation of
taking the partial trace is performed only once. This procedure is no different than that of measuring the spin
of an entangled electron pair: a measurement performed on one of the electrons not only determines that
electron's spin, but also the spin of the second electron. Therefore, even though in a sense the ``refresh''
procedure appears to be nonlinear in $\rho_{12}$, employing it does not break any physical rules of quantum
mechanics.

That being clarified, the ``interact-refresh-repeat'' process of our model is described in more detail below:
\[
\rho_{12}(0)=\rho_1(0)\otimes\rho_2(0)\stackrel{\hat{U}(\tau)}\longrightarrow\rho_{12}(\tau)=\rho_1(\tau)
\otimes\rho_2(\tau)\stackrel{\hat{U}(\tau)}\longrightarrow\ldots\stackrel{\hat{U}(\tau)}
\longrightarrow\rho_{12}(n\tau) =\rho_1(n\tau)\otimes\rho_2(n\tau).
\]
More specifically,
\[
\rho_1(0)\otimes\rho_2(0)\stackrel{\hat{U}(\tau)}\longrightarrow
\mbox{Tr}_2[\hat{U}\rho_{12}(0)\hat{U}^{\dag}]\otimes
\mbox{Tr}_1[\hat{U}\rho_{12}(0)\hat{U}^{\dag}]\stackrel{\hat{U}(\tau)}\longrightarrow\ldots\stackrel{\hat{U}(\tau)}
\longrightarrow \mbox{Tr}_2\left[\hat{U}\rho_{12}[(n-1)\tau]\hat{U}^{\dag}\right]\otimes
\mbox{Tr}_1\left[\hat{U}\rho_{12}[(n-1)\tau]\hat{U}^{\dag}\right],
\]
with the time evolution operator given by:
\[
\hat{U}(\tau)=e^{-i\hat{H}\tau/\hbar}.
\]
A flow chart of our model follows in Fig. \ref{fig:FlowChart}:
\begin{figure}[htp]
\centering
\includegraphics[clip, width=4.6in, height=6.8in, angle=-90]{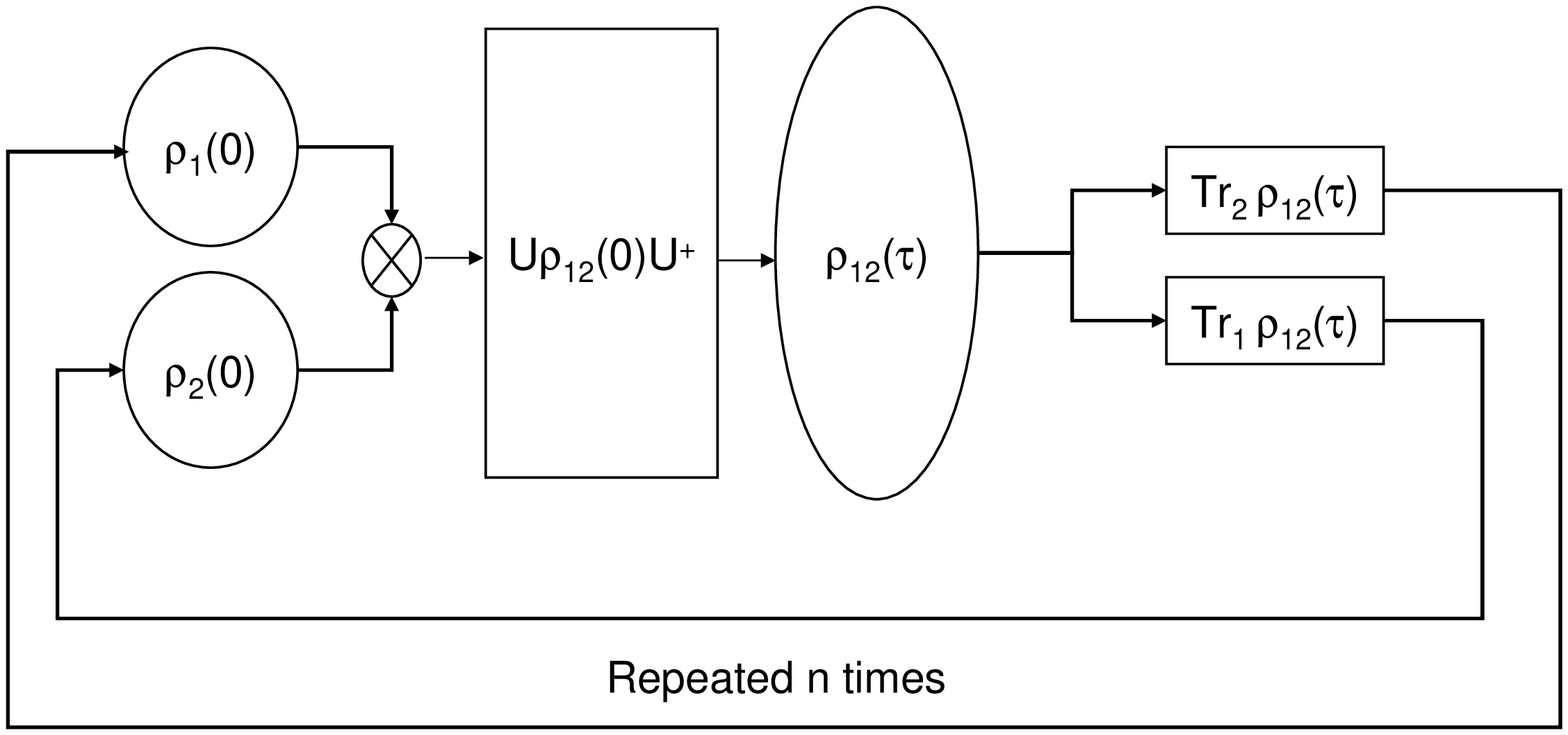}
\caption{Flow Chart describing the first step in the iterative process of ``interacting'', ``refreshing'', and
``repeating''. Two initially uncoupled subsystems are forced to interact through a constant interaction
Hamiltonian $H_{int}$, the time-evolved combined density matrix is calculated, and the system is ``refreshed''.
Statistical independence is assumed at each characteristic time interval $\tau$. The interaction Hamiltonian is
again applied to the new uncoupled states of the two subsystems. The process is periodically repeated until a
new equilibrium is reached.} \label{fig:FlowChart}
\end{figure}

This iteration process is carried out until a new equilibrium is reached. Our aim is to study the large time
behavior of this periodically repeated interaction scheme and understand its effect on the reduced quantum
subsystems $\rho_1$ and $\rho_2$.

The problem of describing the approach to equilibrium for systems composed of a large number of interacting
particles is a very fundamental one, and one that has been studied extensively through the years. Ever since the
classic work of Boltzmann, it has been recognized that not only statistical, but also dynamical considerations
play a role in the time evolution of these systems. Stochastic processes like the one described above are
encountered in many contexts in both physics, chemistry, and, more recently, quantum communication. Relevant
experimental applications involve the investigation of resonance Raman scattering \cite{Shen, Rousseau}, as well
as collisional decoherence during writing and reading quantum states \cite{Manz}.

\section{Evolving Harmonic Oscillator System}
We are now ready to apply the above recipe to the case of two coupled harmonic oscillators. The system
considered here is a set of two initially uncoupled harmonic oscillators $1$ and $2$, described by density
matrices $\rho_1(0)$ and $\rho_2(0)$ respectively. The two subsystems are initially at equilibrium at
temperatures $T_1(0)$ and $T_2(0)$. The free and interaction Hamiltonians of our system are given by:
\begin{eqnarray*}
\hat{H}_{1}&=&\hbar\omega_{1}\hat{a}^\dag_{1}\hat{a}_1\\
\hat{H}_{2}&=&\hbar\omega_{2}\hat{a}^\dag_{2}\hat{a}_2\\
\hat{H}_{int}&=&\hbar\omega\lambda(\hat{a}^\dag_{1}\hat{a}_{2}+\hat{a}^\dag_{2}\hat{a}_{1}),
\end{eqnarray*}
where $\omega_{1,2}$ is the oscillation frequency of oscillator $1,2$, and $\omega$ is the frequency of the
applied interaction Hamiltonian. $\lambda$ is the strength of coupling between the two oscillators. Here,
$\hat{a}^\dag_1$ and $\hat{a}_1$ are the raising and lowering operators for oscillator $1$, and similarly,
$\hat{a}^\dag_2$ and $\hat{a}_2$ are the raising and lowering operators for oscillator $2$. For simplicity, we
will set $\hbar=1$ for the remainder of this paper.

The initial uncoupled and combined density matrices have the Maxwell-Boltzmann distribution and are given by:
\begin{eqnarray}
\rho_{1}(0)&=&\frac{1}{Z_1[\theta_1(0)]}e^{-\omega_{1}\hat{a}^\dag_{1}\hat{a}_{1}\theta_{1}(0)}\nonumber\\
\rho_{2}(0)&=&\frac{1}{Z_2[\theta_2(0)]}e^{-\omega_{2}\hat{a}^\dag_{2}\hat{a}_{2}\theta_{2}(0)}\nonumber\\
\rho_{12}(0)&=&\frac{1}{Z_1[\theta_1(0)]Z_2[\theta_2(0)]}e^{-\omega_{1}\hat{a}^\dag_{1}\hat{a}_{1}\theta_{1}(0)}
e^{-\omega_{2}\hat{a}^\dag_{2}\hat{a}_{2}\theta_{2}(0)}\nonumber\\
&=&\frac{1}{Z_1[\theta_1(0)]Z_2[\theta_2(0)]}e^{-[\omega_{1}\hat{a}^\dag_{1}\hat{a}_{1}\theta_{1}(0)+
\omega_{2}\hat{a}^\dag_{2}\hat{a}_{2}\theta_{2}(0)]}\label{initial},
\end{eqnarray}
where we have used the property $[\hat{a}_1^{\dag}\hat{a}_1,\hat{a}_{2}^{\dag}\hat{a}_2]=0$ to obtain the last
equation. Here, $\theta_{1,2}=1/(kT_{1,2})$, $k$ is the Boltzmann constant and $T_{1,2}$ is the initial
equilibrium temperature of oscillator $1,2$. The quantity $Z$ is a normalization constant, the partition
function, and is given by
$Z_i=\sum{e^{-E_n/(kT)}}=\mbox{Tr}(e^{-\hat{H}_i/(kT)})=\mbox{Tr}(e^{-\omega\hat{a}^\dag_{i}\hat{a}_i\theta_i})$.
The time evolution of the complete system is given by:
\begin{eqnarray}
\rho_{12}(\tau)&=&e^{-i\hat{H}_{total}\tau}\rho_{12}(0)e^{+i\hat{H}_{total}\tau}\nonumber\\
&=&e^{-i(\hat{H}_{1}+\hat{H}_{2}+\hat{H}_{int})\tau}\rho_{12}(0)e^{+i(\hat{H}_{1}+\hat{H}_{2}+
\hat{H}_{int})\tau}\nonumber\\
&=&\frac{1}{Z_1[\theta_1(0)]Z_2[\theta_2(0)]}e^{-i\tau[\omega_1\hat{a}^\dag_{1}\hat{a}_1
+\omega_2\hat{a}^\dag_{2}\hat{a}_2
+\omega\lambda(\hat{a}^\dag_{1}\hat{a}_{2}+\hat{a}^\dag_{2}\hat{a}_{1})]}\nonumber\\
&\times&e^{-[\omega_1\hat{a}^\dag_{1}\hat{a}_{1}\theta_{1}(0)
+\omega_2\hat{a}^\dag_{2}\hat{a}_{2}\theta_{2}(0)]}\nonumber\\
&\times&e^{+i\tau[\omega_1\hat{a}^\dag_{1}\hat{a}_1+\omega_2\hat{a}^\dag_{2}\hat{a}_2+
\omega\lambda(\hat{a}^\dag_{1}\hat{a}_{2}+\hat{a}^\dag_{2}\hat{a}_{1})]}. \label{rho evolved}
\end{eqnarray}
We will eventually extract the reduced evolution of one part by taking the partial trace with respect to the
other part, but first we need to calculate $\rho_{12}(\tau)$ explicitly. To proceed, we note that there is a
very interesting connection between the algebra of angular momentum and the algebra of two uncoupled harmonic
oscillators. This connection is the well-known Schwinger's Oscillator Model of Angular Momentum
\cite{Schwinger}. In this model, we define the following operators:
\begin{eqnarray*}
\hat{J}_{+}&=&\hat{a}_{1}^\dag\hat{a}_{2}\\
\hat{J}_{-}&=&\hat{a}_{2}^\dag\hat{a}_{1}\\
\hat{J}_{1}&=&\frac{1}{2}(\hat{a}_{1}^\dag\hat{a}_{2}+\hat{a}_{2}^\dag\hat{a}_{1})\\
\hat{J}_{2}&=&\frac{1}{2i}(\hat{a}_{1}^\dag\hat{a}_{2}-\hat{a}_{2}^\dag\hat{a}_{1})\\
\hat{J}_{3}&=&\frac{1}{2}(\hat{a}_{1}^\dag\hat{a}_{1}-\hat{a}_{2}^\dag\hat{a}_{2})\\
j&=&\frac{1}{2}(\hat{a}_{1}^\dag\hat{a}_{1}+\hat{a}_{2}^\dag\hat{a}_{2})=\frac{1}{2}(n_{1}+n_{2})\\
m&=&\frac{1}{2}(\hat{a}_{1}^\dag\hat{a}_{1}-\hat{a}_{2}^\dag\hat{a}_{2})=\frac{1}{2}(n_{1}-n_{2}) =\hat{J}_{3}.
\end{eqnarray*}
It can be shown that the above operators satisfy angular momentum commutation relations. With the above
definitions, equation (\ref{rho evolved}) becomes:
\begin{eqnarray*}
\rho_{12}(\tau)&=&\frac{1}{Z_1[\theta_1(0)]Z_2[\theta_2(0)]}e^{-i[j(\omega_{1}+\omega_{2})
+\hat{J}_{3}(\omega_{1}
-\omega_{2})+2\omega\lambda \hat{J}_{1}]\tau}\\
&\times&e^{-{j[\omega_{1}\theta_{1}(0)+\omega_{2}\theta_{2}(0)]+\hat{J}_{3}[\omega_{1}\theta_{1}(0)
-\omega_{2}\theta_{2}(0)]}}\\
&\times&e^{+i[j(\omega_{1}+\omega_{2})+\hat{J}_{3}(\omega_{1}-\omega_{2})+2\omega\lambda \hat{J}_{1}]\tau}.
\end{eqnarray*}
Since $j$ commutes with $\hat{J}_{1}$ and $\hat{J}_{3}$, we can pull it through and factor it out, obtaining a
simplified expression for $\rho_{12}(\tau)$:
\begin{eqnarray}
\rho_{12}(\tau)&=&\frac{1}{Z_1[\theta_1(0)]Z_2[\theta_2(0)]}e^{-{j[\omega_{1}\theta_{1}(0)
+\omega_{2}\theta_{2}(0)]}}
\nonumber\\
&\times&e^{-i\tau[\hat{J}_{3}(\omega_{1}-\omega_{2})+2\omega\lambda \hat{J}_{1}]}\nonumber\\
&\times&e^{-\hat{J}_{3}[\omega_{1}\theta_{1}(0)-\omega_{2}\theta_{2}(0)]}\nonumber\\
&\times&e^{+i\tau[\hat{J}_{3}(\omega_{1}-\omega_{2})+2\omega\lambda \hat{J}_{1}]}.\label{rho}
\end{eqnarray}

By making the following definitions:
\begin{eqnarray}
a&=&\tau(\omega_{1}-\omega_{2})\nonumber\\
b&=&2\omega\lambda \tau\nonumber\\
c&=&i[\omega_{2}\theta_{2}(0)-\omega_{1}\theta_{1}(0)], \label{abc}
\end{eqnarray}
we can rewrite equation (\ref{rho}) into its final form as:
\begin{equation}
\rho_{12}(\tau)=\frac{1}{Z_1[\theta_1(0)]Z_2[\theta_2(0)]}e^{-j[\omega_1\theta_1(0)+\omega_2\theta_2(0)]}
e^{-i(a\hat{J}_3+b\hat{J}_1)}e^{-ic\hat{J}_3}e^{+i(a\hat{J}_3+b\hat{J}_1)}.\label{finalform}
\end{equation}

The reduced time evolution of oscillator $1$ after a time $\tau$ is obtained by taking the partial trace of
equation (\ref{finalform}) over oscillator $2$. The matrix elements of such an expression are given by:
\begin{eqnarray}
\langle n_1'|\rho_1(\tau)|n_1\rangle&=&\sum_{n_2}\langle n_1'n_2|\rho_{12}(\tau)|n_1n_2 \rangle
=\delta_{n_1'n_1}\sum_{n_2}\langle n_1'n_2|\rho_{12}(\tau)|n_1n_2 \rangle\nonumber\\
&=&\frac{1}{Z_1[\theta_1(0)]Z_2[\theta_2(0)]}\sum_{n_2}\langle
n_1n_2|e^{-j[\omega_1\theta_1(0)+\omega_2\theta_2(0)]}\nonumber\\
&\times&e^{-i(a\hat{J}_3+b\hat{J}_1)}\times e^{-ic\hat{J}_3}\times e^{+i(a\hat{J}_3+b\hat{J}_1)}|n_1n_2
\rangle,\label{matrix elements}
\end{eqnarray}

As promised in section II, we now discuss the physical realizations of the ``refresh'' procedure. To begin, we
note that the reduced density matrix $\rho_1(\tau)$ is diagonal in the number basis $\{|n_1\rangle\}$ of the
first harmonic oscillator. This is due to the fact that
$j=N=1/2(\hat{a}_1^{\dag}\hat{a}_1+\hat{a}_2^{\dag}\hat{a}_2)$ is a conserved quantity in our model $([H,N]=0)$,
and also because the initial density matrix $\rho_{12}(0)$ is diagonal in the number basis $\{|n_1n_2\rangle\}$
of the harmonic oscillators. The evolution generated by the interaction Hamiltonian cannot change the value of
the conserved quantity. Since the only non-zero elements in $\rho_{12}(0)$ are those satisfying
$n_1+n_2=n_1'+n_2'$, the only non-zero matrix elements of the time-evolved density matrix $\rho_{12}(\tau)$ are
$\langle n_1n_2|\rho_{12}(\tau)|n_1'n_2'\rangle$ with $n_1+n_2=n_1'+n_2'$. In performing the partial trace over
the second harmonic oscillator, we set $n_2=n_2'$, which means that all the terms of $\rho_{12}(\tau)$
contributing to the partial trace must also have $n_1=n_1'$. Like mentioned in Sec. II, the operation of taking
the partial trace over the second oscillator is equivalent to making a \emph{non-selective}, or spectrum
measurement in the number basis of the first oscillator. It so happens that because of the conservation of the
global number of particles, this one spectrum measurement on subsystem $1$, not only determines the probability
distribution of the occupations numbers of the first harmonic oscillator, but it also automatically determines
that of the second oscillator. In addition, it simultaneously diagonalizes the output state in both number bases
$\{|n_1\rangle\}$ and $\{|n_2\rangle\}$ of the two oscillators. Since the output state is now diagonal in both
number bases $\{|n_1\rangle\}$ and $\{|n_2\rangle\}$, it can be written as the tensor product of the two reduced
post-measurement subsystem states and statistical independence between the two subsystems can be assumed. Even
though the ``refresh'' procedure is expressed as the tensor product of the two post-measurement states, obtained
by taking partial traces over each unwanted part, the actual operation of taking the partial trace is performed
only once, and over only one of the subsystems. A single measurement determines the outcome of not just one, but
both oscillator systems.

We now need to figure out how the exponential operators in equation (\ref{matrix elements}) act on our
two-system state $|n_1n_2\rangle$. This calculation is greatly simplified if we make a transformation to the
Euler-angle form $e^{-i\alpha \hat{J}_{3}}e^{-i\beta \hat{J}_{2}}e^{-i\gamma \hat{J}_{3}}$, since the action of
the operators $\hat{J}_3$ and $\hat{J}_2$ on our state is a well-known result. We would like to find expressions
for the coefficients $\alpha$, $\beta$, and $\gamma$ that appear in the Euler angle form, in terms of the
coefficients $a$, $b$, and $c$ given in equation (\ref{abc}). To do this, we make use of the well-known
identity:
\[
e^{-i\hat{n}\cdot\vec{\sigma}\frac{\phi}{2}}=I\cos\left(\frac{\phi}{2}\right)-i\hat{n}\cdot\vec{\sigma}\sin
\left(\frac{\phi}{2}\right),
\]
where, for our case,
\[
\vec{\sigma}=\sigma_{1}\hat{i}+\sigma_{2}\hat{j}+\sigma_{3}\hat{k}=2\hat{J},
\]
(the $\sigma_{1,2,3}$ are the Pauli spin matrices) and
\[
\hat{n}=\frac{1}{\sqrt{(\frac{a}{2})^2+(\frac{b}{2})^2}}\left(\frac{b}{2}\hat{i}+\frac{a}{2}\hat{k}\right),
\]
such that
\[
\frac{\phi}{2}=\sqrt{\left(\frac{a}{2}\right)^2+\left(\frac{b}{2}\right)^2}.
\]
After some extensive algebraic calculations, the results are:
\begin{eqnarray*}
\alpha&=&\arctan\left(\frac{D}{A}\right)+\arctan\left(-\frac{B}{C}\right)\\
\beta&=&2\arccos\left[\frac{A}{\cos\left(\arctan{\frac{D}{A}}\right)}\right]\\
\gamma&=&\arctan\left(\frac{D}{A}\right)-\arctan\left(-\frac{B}{C}\right),
\end{eqnarray*}
where
\begin{eqnarray*}
A&=&\cos\left(\frac{c}{2}\right)\\
B&=&\frac{a b}{2d^2}\sin^2(d)\sin\left(\frac{c}{2}\right)\\
C&=&-\frac{b}{d}\cos(d)\sin(d)\sin\left(\frac{c}{2}\right)\\
D&=&\left[\cos^2(d)+\frac{\left(\frac{a}{2}\right)^2-\left(\frac{b}{2}\right)^2}{d^2}\sin^2(d)\right]
\sin\left(\frac{c}{2}\right),
\end{eqnarray*}
and
\[
d=\sqrt{\left(\frac{a}{2}\right)^2+\left(\frac{b}{2}\right)^2}=\frac{\phi}{2}.
\]
With the above transformations, equation (\ref{matrix elements}) becomes:
\begin{eqnarray}
\langle
n_{1}'|\rho_{1}(t)|n_{1}\rangle&=&\frac{\delta_{n_{1}'n_1}}{Z_1[\theta_1(0)]Z_2[\theta_2(0)]}\sum_{n_{2}}
\langle n_{1}'n_{2}|e^{-j[\omega_{1}\theta_{1}(0)+\omega_{2}\theta_{2}(0)]}e^{-i\alpha\hat{J}_{3}}e^{-i\beta
\hat{J}_{2}}e^{-i\gamma \hat{J}_{3}}|n_{1}n_{2}\rangle\nonumber\\
&=&\frac{1}{Z_1[\theta_1(0)]Z_2[\theta_2(0)]}\sum_{n_{2}}e^{-\frac{1}{2}(n_{1}+n_{2})[\omega_{1}\theta_{1}(0)
+\omega_{2}\theta_{2}(0)]}e^{-\frac{i}{2}(n_{1}-n_{2})(\alpha+
\gamma)}\nonumber\\
& &\times\langle n_{1}n_{2}|e^{-i\beta \hat{J}_{2}}|n_{1}n_{2}\rangle,\label{combined}
\end{eqnarray}
where the $\hat{J}_2$ matrix element in the last equation is given by Wigner's Formula in the $|j,m\rangle$
basis \cite{Sakurai}:
\begin{eqnarray*}
\langle j,m'|e^{-i\beta \hat{J}_{2}}|j,m\rangle&=&
d_{m',m}^{(j)}(\beta)\nonumber\\
&=&\sum_{K}(-1)^{K}\frac{\sqrt{(j+m)!(j-m)!(j+m')!(j-m')!}}{(j-m'-K)!K!(j+m-K)!(K+m'-m)!}\nonumber\\
&\times&\left(\cos\frac{\beta}{2}\right)^{j+m+j-m'-2K}\left(\sin\frac{\beta}{2}\right)^{2K+m-m'}.
\end{eqnarray*}
Substituting this in equation (\ref{combined}) and summing over $n_2$ and $K$, we obtain:
\begin{eqnarray}
\langle n_1'|\rho_1(\tau)|n_1\rangle&=&\frac{\delta_{n_1'n_1}}{Z_1[\theta_1(0)]Z_2[\theta_2(0)]}\times
\left(\frac{1}{1-\cos\left(\frac{\beta}{2}\right)e^{\frac{i}{2}(\alpha+\gamma)-\Theta}}\right)\nonumber\\
&\times&\left[\left(\frac{\cos\left(\frac{\beta}{2}\right)}{e^{\Theta+\frac{i}{2}(\alpha+\gamma)}}
\right)\left(\frac{1-\sec\left(\frac{\beta}{2}\right)e^{\frac{i}{2}(\alpha+\gamma)-\Theta}}{1-\cos
\left(\frac{\beta}{2}\right)e^{\frac{i}{2}(\alpha+\gamma)-\Theta}}\right)\right]^{n_1}, \label{firstrau}
\end{eqnarray}
where
\[
\Theta=\frac{1}{2}[\omega_1\theta_1(0)+\omega_2\theta_2(0)].
\]
Defining $\alpha=i\zeta$ and $\gamma=i\xi$ in equation (\ref{firstrau}) and letting
$\delta=\frac{1}{2}(\zeta+\xi)$, we finally get:
\begin{eqnarray}
\langle n_1'|\rho_1(\tau)|n_1\rangle&=&\frac{\delta_{n_1'n_1}}{Z_1[\theta_1(0)]Z_2[\theta_2(0)]}
\left[\frac{\cos(\frac{\beta}{2})e^{\delta+\Theta}-1}{e^{\Theta-\delta}
[e^{\Theta+\delta}-\cos(\frac{\beta}{2})]}\right]^{n_1}\nonumber\\
&\times&\frac{1}{1-\cos(\frac{\beta}{2})e^{-(\Theta+\delta)}}.\label{secondrau}
\end{eqnarray}
Equation (\ref{secondrau}) can be written in the form:
\begin{equation}
\langle
n_1'|\rho_1(\tau)|n_1\rangle=\delta_{n_1'n_1}\frac{1}{Z[\theta_1(\tau)]}e^{-\omega_1n_1\theta_1(\tau)},\label{Boltzmann}
\end{equation}
with
\[
e^{-\omega_1\theta_1(\tau)}=\frac{\cos(\frac{\beta}{2})e^{\Theta+\delta}-1}{e^{\Theta-\delta}
[e^{\Theta+\delta}-\cos(\frac{\beta}{2})]}
\]
and
\[
Z[\theta_1(\tau)]=Z_1[\theta_1(0)]Z_2[\theta_2(0)]\left[1-\cos(\frac{\beta}{2})e^{-(\Theta+\delta)}\right].
\]
We could have obtained the matrix elements of the reduced density matrix of oscillator $2$, by taking the
partial trace of the time-evolved density matrix of the complete system over oscillator $1$, by following a
similar procedure as the one presented one.

Equation (\ref{Boltzmann}) is a quasi-stationary, non-equilibrium state. It states that, after the interaction
Hamiltonian has been applied for a time interval $\tau$, the original Maxwell-Boltzmann distribution describing
oscillator $1$ with temperature $T_1(0)=1/[k\theta_1(0)]$ at $t=0$, is replaced by another Maxwell-Boltzmann
distribution and a different temperature $T_1(\tau)=1/[k\theta_1(\tau)]$. More specifically, the interaction has
forced the oscillator out of equilibrium at temperature $T_1(0)$ and brought it to a new equilibrium at
temperature $T_1(\tau)$. A similar statement is true of oscillator $2$.

Now that we have the matrix elements of the reduced time-evolved density matrix of oscillator $1$ at time
$\tau$, we can repeat our calculations to obtain $\langle n_1'|\rho_1(2\tau)|n_1\rangle$, which will lead to
another Maxwell-Boltzmann distribution (and another quasi-stationary, non-equilibrium state), with a different
temperature $T_1(2\tau)=1/[k\theta_1(2\tau)]$. The above process can be repeated to obtain $\langle
n_1'|\rho_1(3\tau)|n_1\rangle$, $\langle n_1'|\rho_1(4\tau)|n_1\rangle$, ...., $\langle
n_1'|\rho_1(n\tau)|n_1\rangle$ with an increasing level of difficulty.

\section{Physical Interpretation and Consequences}
The question we want to answer now is: \emph{When is equilibrium achieved}? To answer this question, let's
consider the system composed of some energy state $|n_1\rangle$ in oscillator $1$ and some energy state
$|n_2\rangle$ in oscillator $2$. Equilibrium will be reached when the net rate of ``absorption'' equals the net
rate of ``emission'' in and out of our system. ``Emission'' results from transitions such as
$|n_1\rangle\rightarrow |n_1+1\rangle$ or $|n_1\rangle\rightarrow |n_1-1\rangle$ in oscillator $1$, and
$|n_2\rangle\rightarrow |n_2-1\rangle$ or $|n_2\rangle\rightarrow |n_2+1\rangle$ in oscillator $2$. Likewise,
``absorption'' processes are caused from transitions such as $|n_1-1\rangle\rightarrow |n_1\rangle$ or
$|n_1+1\rangle\rightarrow |n_1\rangle$ and $|n_2+1\rangle\rightarrow |n_2\rangle$ or $|n_2-1\rangle\rightarrow
|n_2\rangle$ for $|n_2\rangle$.

Recalling that $\hat{H}_{int}=\omega\lambda(\hat{a}^\dag_{1}\hat{a}_{2}+\hat{a}^\dag_{2}\hat{a}_{1})$, we can
calculate the probability amplitudes for ``emission'' out of this bipartite system:
\[
\langle n_1+1,n_2-1|\omega\lambda(\hat{a}^\dag_{1}\hat{a}_{2}+\hat{a}^\dag_{2}\hat{a}_{1})|n_1,n_2\rangle
=\omega\lambda\sqrt{n_2(n_1+1)},
\]
and
\[
\langle n_1-1,n_2+1|\omega\lambda(\hat{a}^\dag_{1}\hat{a}_{2}+\hat{a}^\dag_{2}\hat{a}_{1})|n_1,n_2\rangle
=\omega\lambda\sqrt{n_1(n_2+1)}.
\]
Similarly, the probability amplitudes for ``absorption'' are given by:
\[
\langle n_1,n_2|\omega\lambda(\hat{a}^\dag_{1}\hat{a}_{2}+\hat{a}^\dag_{2}\hat{a}_{1})|n_1-1,n_2+1\rangle
=\omega\lambda\sqrt{n_1(n_2+1)},
\]
and
\[
\langle n_1,n_2|\omega\lambda(\hat{a}^\dag_{1}\hat{a}_{2}+\hat{a}^\dag_{2}\hat{a}_{1})|n_1+1,n_2-1\rangle
=\omega\lambda\sqrt{n_2(n_1+1)}.
\]
At equilibrium, we expect that:
\begin{equation}
P(n_1,n_2)[n_2(n_1+1)]+P(n_1,n_2)[n_1(n_2+1)]=
P(n_1-1,n_2+1)[n_1(n_2+1)]+P(n_1+1,n_2-1)[n_2(n_1+1)],\label{probability}
\end{equation}
where $P(n_1,n_2)$ is the probability to have a particle in energy level $|n_1\rangle$ in oscillator $1$ and
energy level $|n_2\rangle$ in oscillator $2$, etc.

Simplifying equation (\ref{probability}) gives:
\begin{eqnarray}
P(n_1,n_2)[2n_1n_2+n_1+n_2]&=&P(n_1-1,n_2+1)[n_1(n_2+1)]\nonumber\\
&+&P(n_1+1,n_2-1)[n_2(n_1+1)]\label{general}.
\end{eqnarray}

We can now explore the consequences of equation (\ref{general}). First, we note that the interaction Hamiltonian
imposes the condition that the sum of the occupation numbers $n_1$ and $n_2$ in states $|n_1\rangle$ and
$|n_2\rangle$ respectively, is a constant. Let's call this constant $N$, such that $n_1+n_2=N$. The case $N=0$
is trivial, it represents the vacuum state. For $N=1$, there are two possible transition states, $|1,0\rangle$
and $|0,1\rangle$. Solving equation (\ref{general}) for these two different cases leads to the condition
$P(1,0)=P(0,1)$. This result is not surprising. It simply states that, at equilibrium, the probability of a
transition to the state $|1,0\rangle$ is the same as that to the state $|0,1\rangle$. Similarly, for $N=2$,
there are three possible transition states: $|0,2\rangle$, $|1,1\rangle$, and $|2,0\rangle$. This time, equation
(\ref{general}) yields a more interesting result: $P(0,2)=P(1,1)=P(2,0)$. Since the sum of all probabilities
must be unity, each of these must be $1/3$. Repeating this procedure for increasing $N$, we soon recognize that
for any $N$, there are $N+1$ different possible states, and the probability of transition to any of them is the
same for all and equal to $1/(N+1)$.

Making use of the above result and employing equations (\ref{initial}) at $t=0$ and (\ref{Boltzmann}) at
equilibrium, we conclude that the new density matrix for the combined system will be of the form:
\begin{equation}
\langle n_1,n_2|\rho_{12}(\infty)|n_1,n_2\rangle=\frac{1}{N+1}e^{-\omega_1n_1\theta_1(\infty)}
e^{-\omega_2n_2\theta_2(\infty)}.\label{ateq}
\end{equation}
Recalling that $N=n_1+n_2$ and defining $\nu=n_1-n_2$, we can rewrite equation (\ref{ateq}) as:
\begin{eqnarray}
\langle
\frac{N+\nu}{2},\frac{N-\nu}{2}|&\rho_{12}(\infty)&|\frac{N+\nu}{2},\frac{N-\nu}{2}\rangle\nonumber\\
&=&\frac{1}{N+1}e^{-\omega_1(\frac{N+\nu}{2})\theta_1(\infty)}e^{-\omega_2(\frac{N-\nu}{2})\theta_2(\infty)}
\nonumber\\
&=&\frac{1}{N+1}\left[e^{-\frac{N}{2}[\omega_1\theta_1(\infty)+\omega_2\theta_2(\infty)]}
e^{-\frac{\nu}{2}[\omega_1\theta_1(\infty)-\omega_2\theta_2(\infty)]}\right].\nonumber\\
\label{Nandnu}
\end{eqnarray}
Since, for any $N$, the probability of transition at equilibrium to any of the $N+1$ possible states is equal to
$1/(N+1)$, equation (\ref{Nandnu}) should be independent of $\nu$. This is true when
$e^{-\frac{\nu}{2}[\omega_1\theta_1(\infty)-\omega_2\theta_2(\infty)]}\rightarrow 1$, which leads to the
consequence that, at equilibrium, $\omega_1\theta_1(\infty)=\omega_2\theta_2(\infty)$, or
$\omega_1/[kT_1(\infty)]=\omega_2/[kT_2(\infty)]$. Note that this result is irrespective of the initial
temperature of the oscillators.

\begin{figure}[htp]
\centering \subfigure[]{\label{fig:LargeTauPrime}\includegraphics[clip, width=3.0in,
height=2.0in]{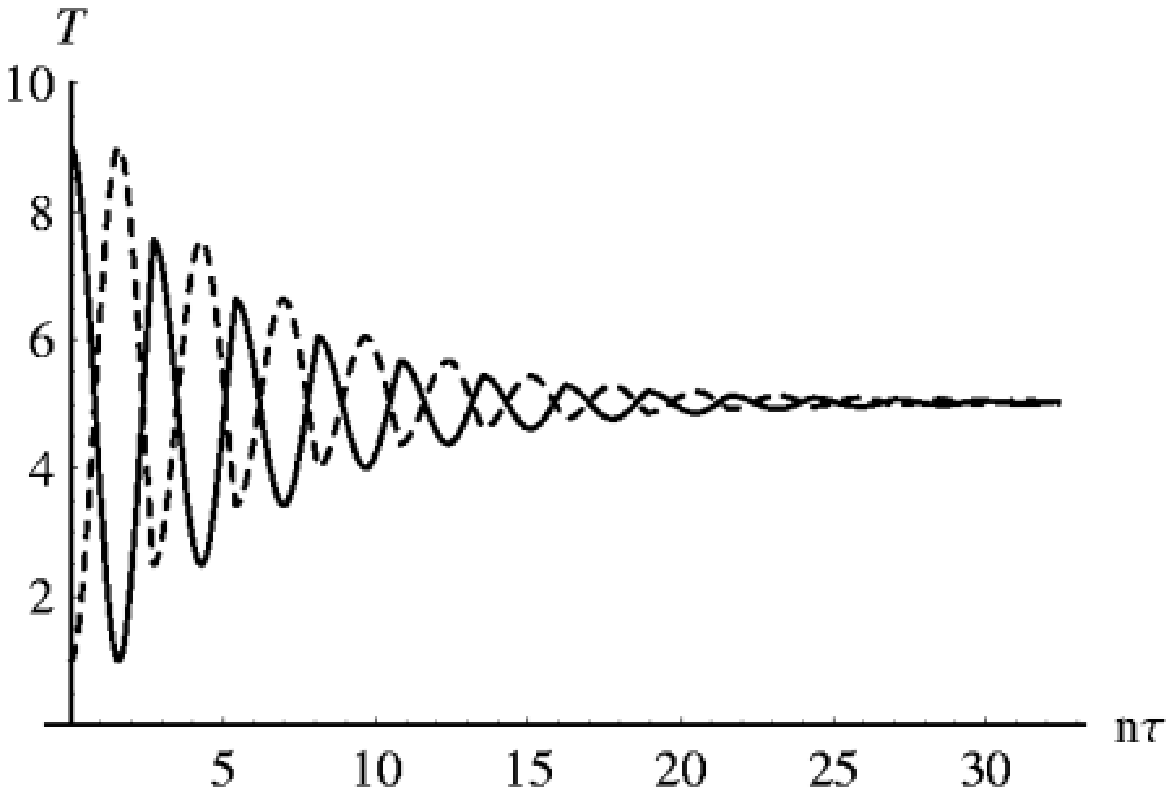}} \subfigure[]{\label{fig:LargeTau}\includegraphics[clip, width=3.0in,
height=2.0in]{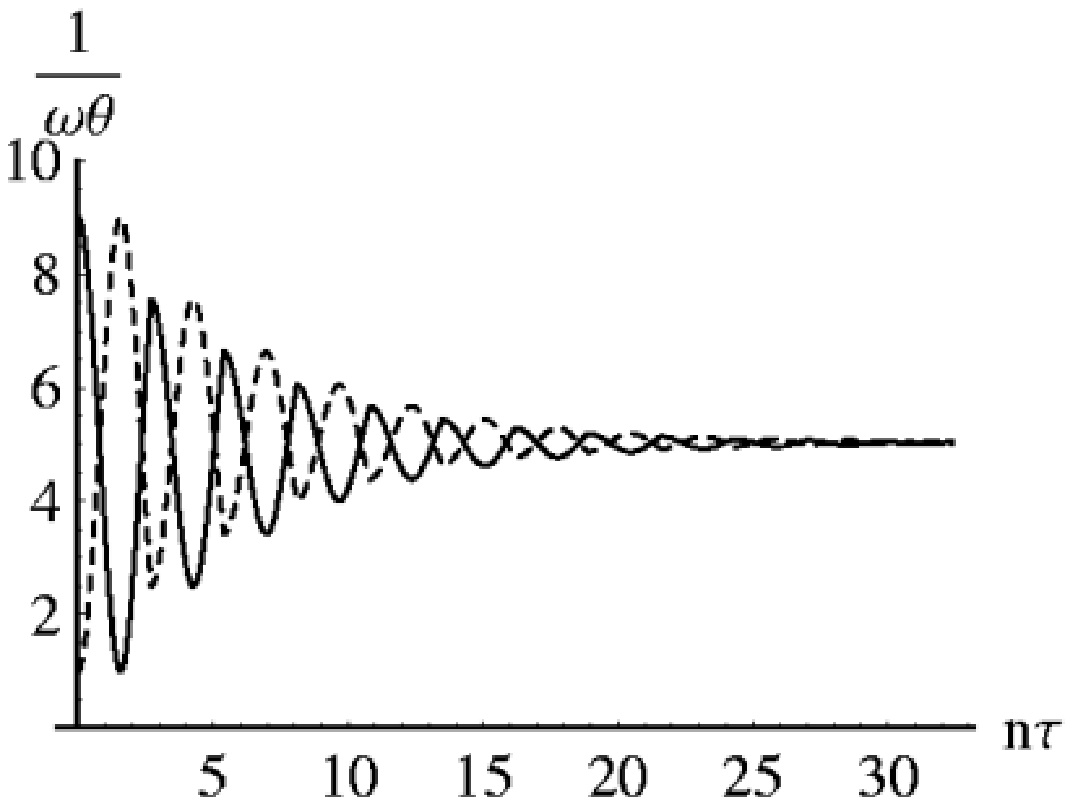}} \caption{$(kt_0/\hbar)T$ and $kT/\hbar\omega$ as functions of the refreshing
time intervals $n\tau/t_0$ for the relaxation of the two initial Maxwell-Boltzmann distributions. The dashed
line represents oscillator $1$ and the solid one represents oscillator $2$. Here, $k=\hbar=t_0=1$,
$(kt_0/\hbar)T_1(0)=1$, $(kt_0/\hbar)T_2(0)=9$, $\omega_1t_0=\omega_2t_0=\omega t_0=1$, and $\tau/t_0=2.7$. The
frequencies of the oscillators are the same and the equilibrium reached is a thermal one.}
\label{fig:FigureLargeTau}
\end{figure}

\begin{figure}[htp]
\centering \subfigure[]{\label{fig:SmallTauPrime}\includegraphics[clip, width=3.0in,
height=2.0in]{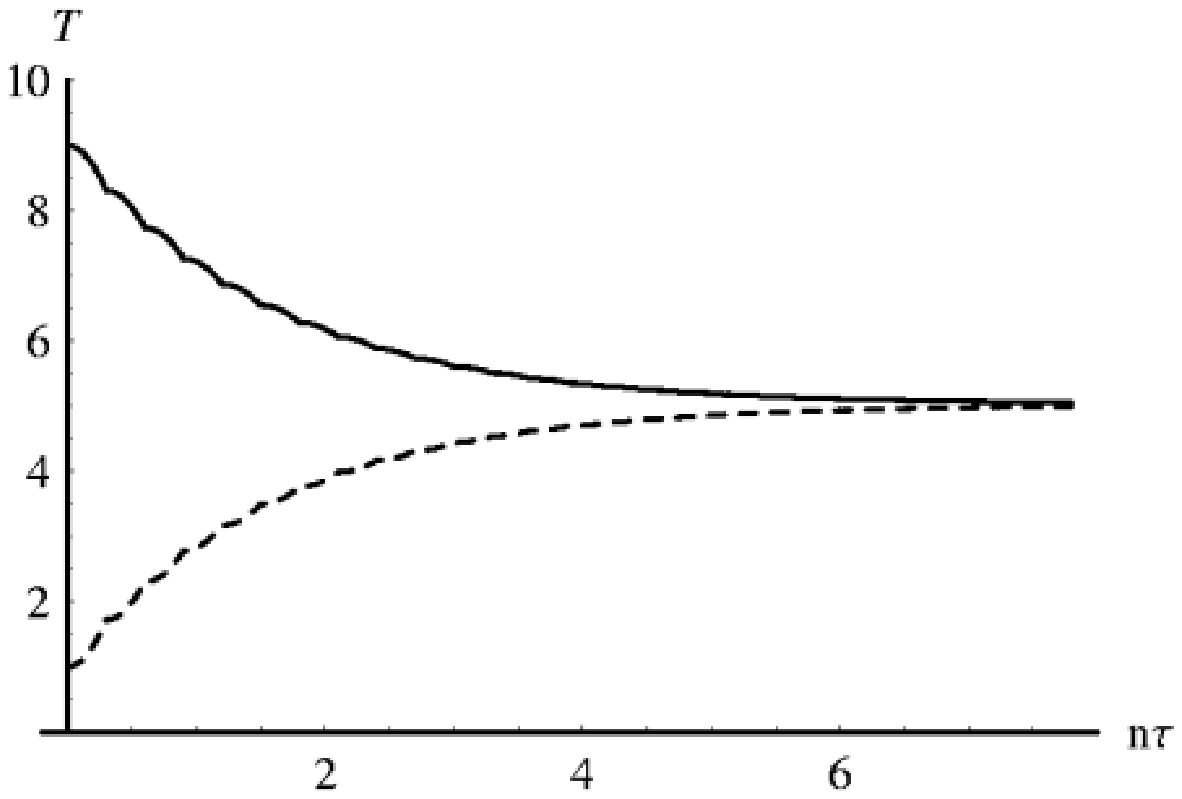}} \subfigure[]{\label{fig:SmallTau}\includegraphics[clip, width=3.0in,
height=2.0in]{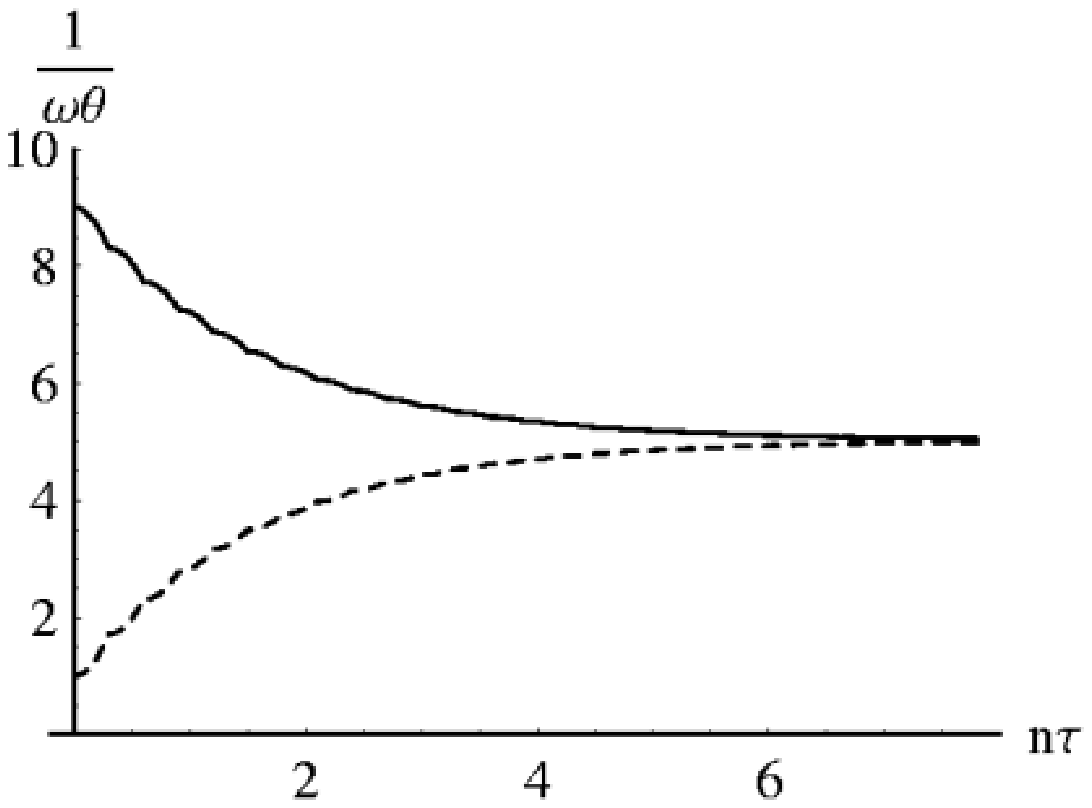}} \caption{$(kt_0/\hbar)T$ and $kT/\hbar\omega$ as functions of the refreshing
time intervals $n\tau/t_0$ for the relaxation of the two initial Maxwell-Boltzmann distributions. The dashed
line represents oscillator $1$ and the solid one represents oscillator $2$. Here, $k=\hbar=t_0=1$,
$(kt_0/\hbar)T_1(0)=1$, $(kt_0/\hbar)T_2(0)=9$, $\omega_1t_0=\omega_2t_0=\omega t_0=1$, and $\tau/t_0=0.3$.
Keeping the oscillation frequencies the same as in Fig. \ref{fig:FigureLargeTau}, the characteristic time
interval is decreased and the equilibrium is reached much faster.} \label{fig:FigureSmallTau}
\end{figure}

\begin{figure}[htp]
\centering \subfigure[]{\label{fig:HighFrequencyPrime}\includegraphics[clip, width=3.0in, height=2.0in]
{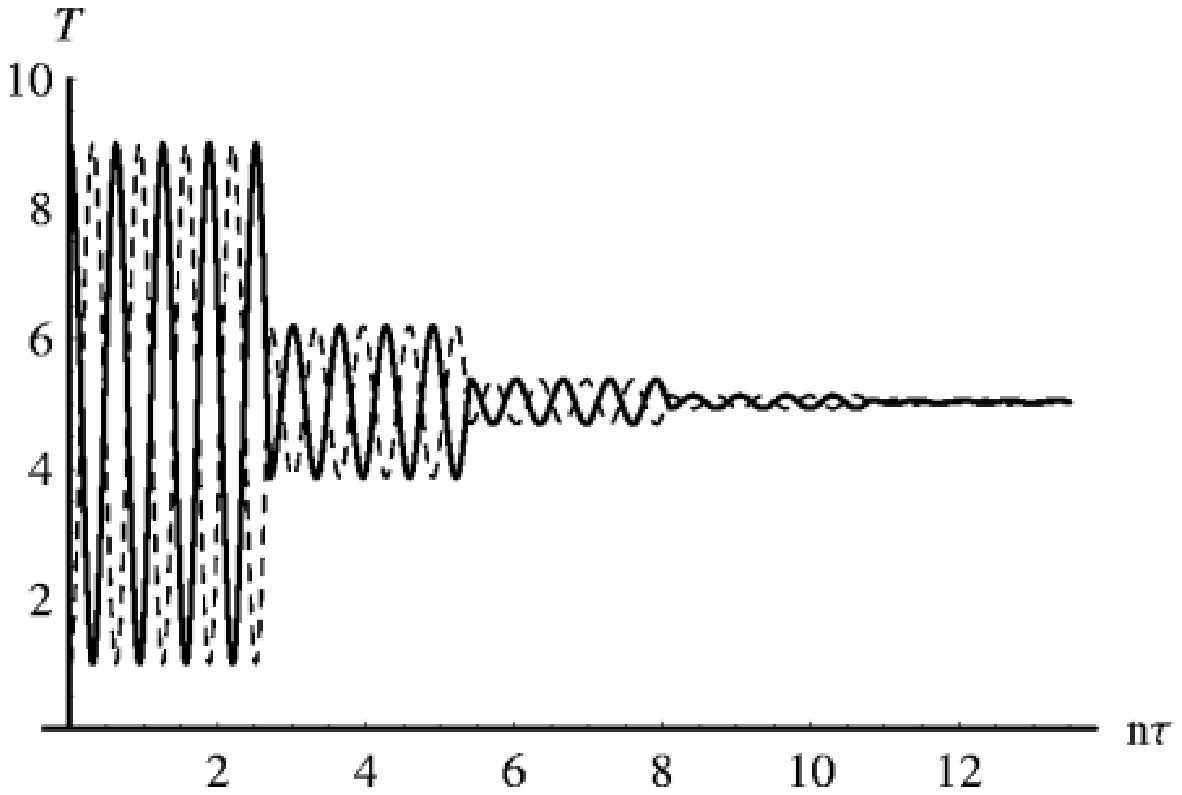}} \subfigure[]{\label{fig:HighFrequency}\includegraphics[clip, width=3.0in,
height=2.0in]{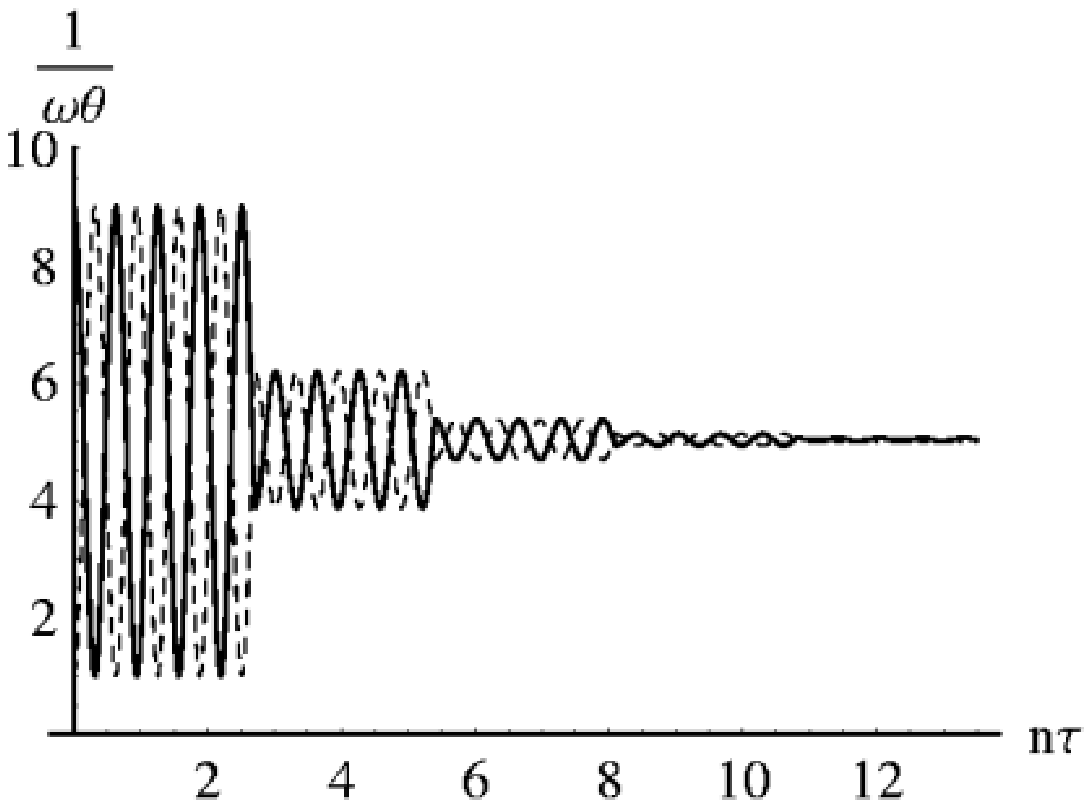}} \caption{$(kt_0/\hbar)T$ and $kT/\hbar\omega$ as functions of the
refreshing time intervals $n\tau/t_0$ for the relaxation of the two initial Maxwell-Boltzmann distributions. The
dashed line represents oscillator $1$ and the solid one represents oscillator $2$. Here, $k=\hbar=t_0=1$,
$(kt_0/\hbar)T_1(0)=1$, $(kt_0/\hbar)T_2(0)=9$, $\omega_1t_0=\omega_2t_0=1$, $\omega t_0=5$, and $\tau/t_0=2.7$.
Keeping the characteristic time interval the same as in Fig. \ref{fig:FigureLargeTau}, the interaction frequency
is increased and the equilibrium is reached much faster.} \label{fig:FigureHighFrequency}
\end{figure}

In the plots below, the ``temperature'' $(kt_0/\hbar)T$ and the quantity $kT/\hbar\omega$ are plotted as
functions of the refreshing time intervals $n\tau/t_0$, for the relaxation of the initial Maxwell-Boltzmann
distributions of the two oscillators. Here, we introduce the time constant $t_0$ for the simplicity of working
in a dimensional system of units. In this system, we set $k=\hbar=t_0=1$. The interaction Hamiltonian is applied
for a constant time interval $\tau$, interrupted, the system is refreshed, and the procedure is repeated, until
a new equilibrium is reached. In Figs. \ref{fig:FigureLargeTau} and \ref{fig:FigureSmallTau}, the frequencies of
both harmonic oscillators and that of the interaction Hamiltonian are equal, in Fig.
\ref{fig:FigureHighFrequency}, the frequencies of the oscillators are equal, but that of the interaction
Hamiltonian is not, and, finally, in Figs. \ref{fig:Figure2} and \ref{fig:Figure4}, the frequencies of the
oscillators and the interaction Hamiltonian are all different. The dashed line represents oscillator $1$ and the
solid one represents oscillator $2$.

\begin{figure}[htp]
\centering \subfigure[]{\label{fig:Figure2Prime}\includegraphics[clip, width=3.0in,
height=2.0in]{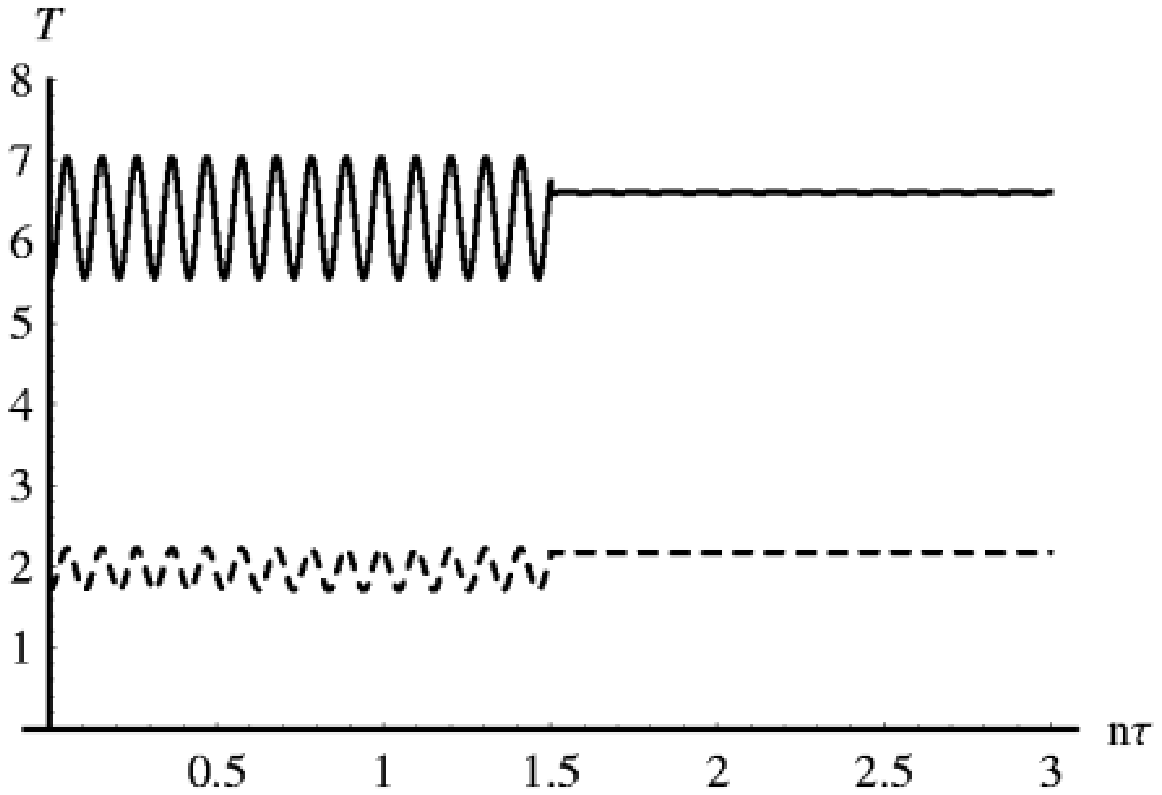}} \subfigure[]{\label{fig:Figure2}\includegraphics[clip, width=3.0in,
height=2.0in]{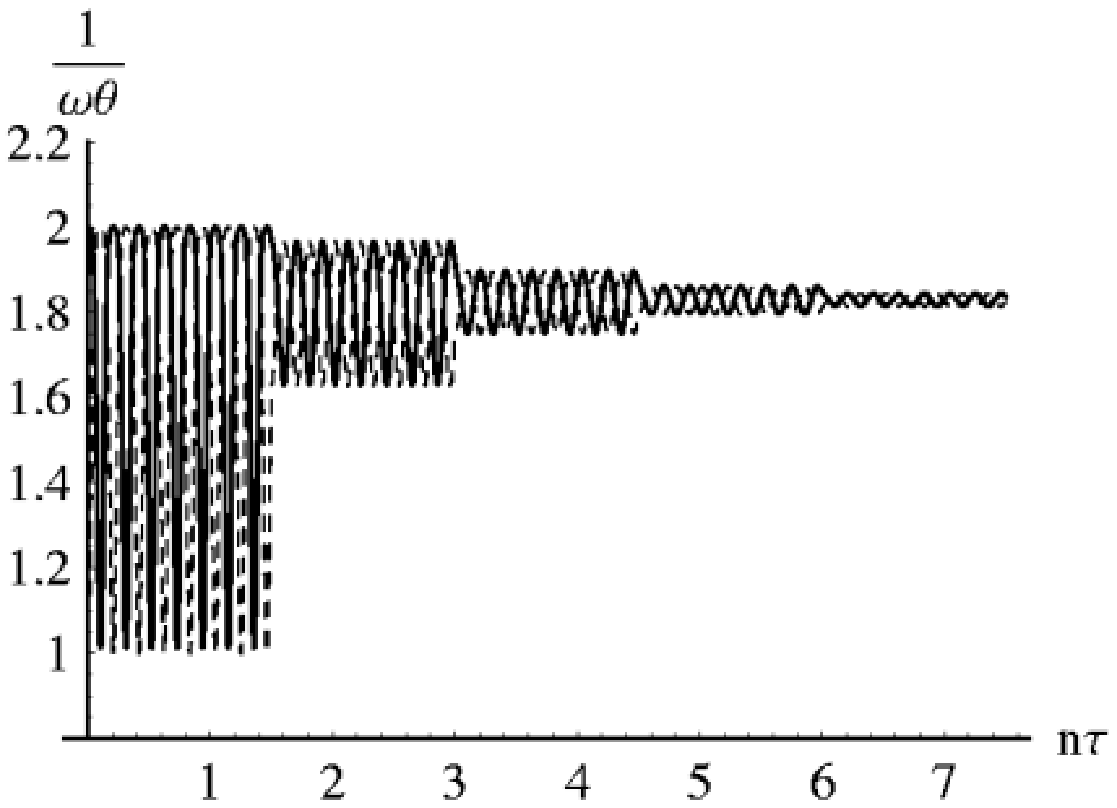}} \caption{$(kt_0/\hbar)T$ and $kT/\hbar\omega$ as functions of the refreshing time
intervals $n\tau/t_0$ for the relaxation of the two initial Maxwell-Boltzmann distributions. The dashed line
represents oscillator $1$ and the solid one represents oscillator $2$. Here, $k=\hbar=t_0=1$,
$(kt_0/\hbar)T_1(0)=2$, $(kt_0/\hbar)T_2(0)=6$, $\omega_1t_0=1$, $\omega_2t_0=3$, $\omega t_0=5$, and
$\tau/t_0=1.5$. The oscillator frequencies are unequal and the equilibrium reached is not thermal. Figure (a)
shows that no single temperature can be assigned to the new equilibrium distribution, and Fig.(b) demonstrates
that the condition $\omega_1/[kT_1(\infty)]=\omega_2/[kT_2(\infty)]$ is satisfied.} \label{fig:Figure2}
\end{figure}

\begin{figure}[htp]
\centering \subfigure[]{\label{fig:Figure4Prime}\includegraphics[clip, width=3.0in,
height=2.0in]{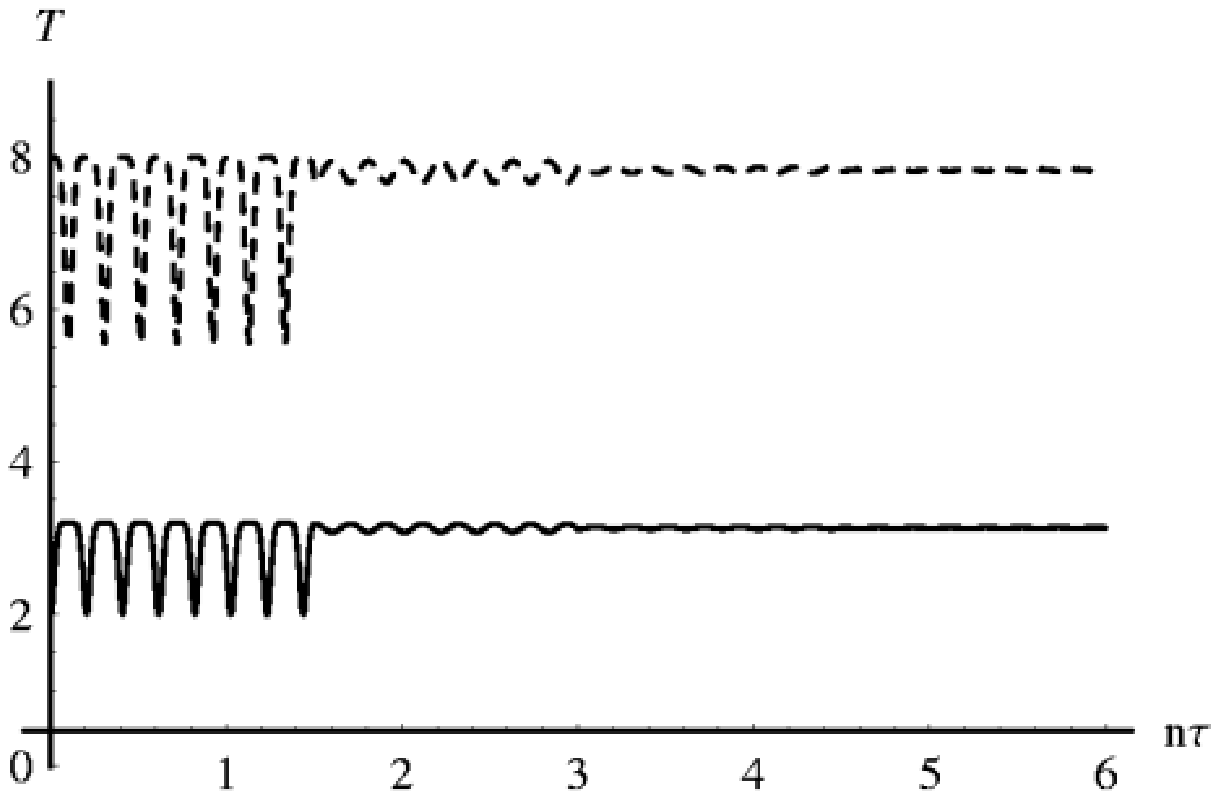}} \subfigure[]{\label{fig:Figure4}\includegraphics[clip, width=3.0in,
height=2.0in]{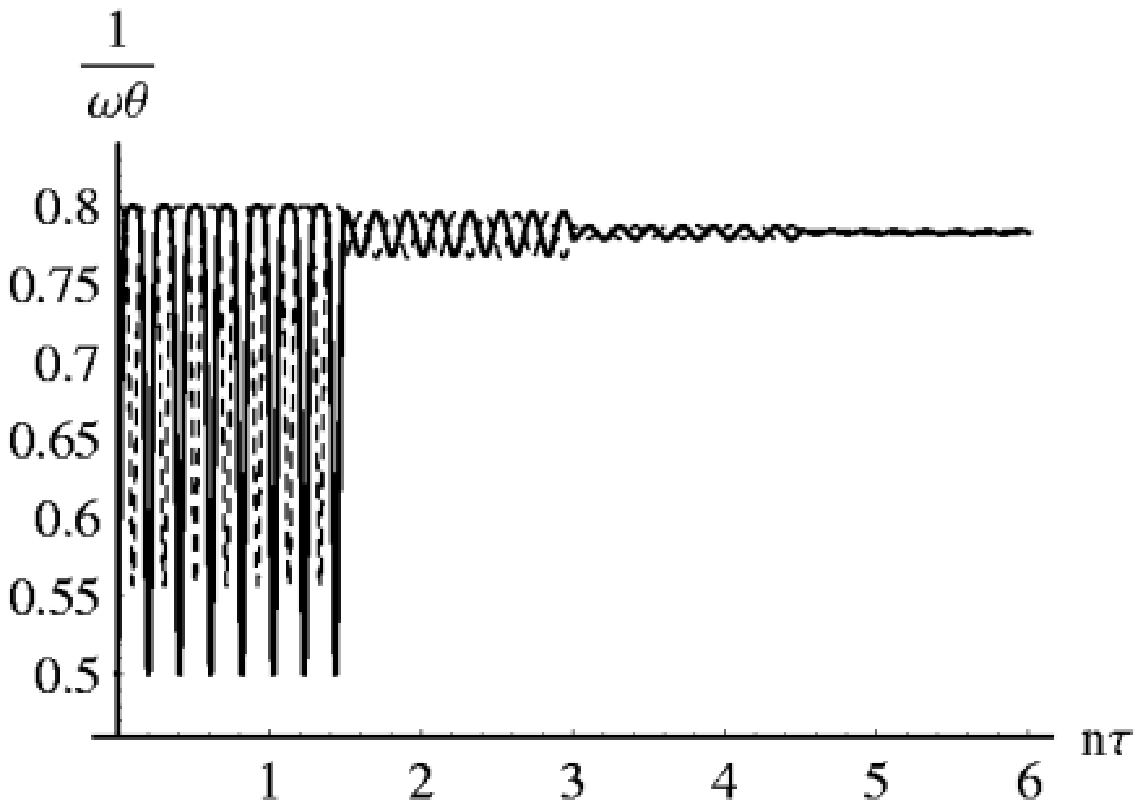}} \caption{$(kt_0/\hbar)T$ and $kT/\hbar\omega$ as functions of the refreshing time
intervals $n\tau/t_0$ for the relaxation of the two initial Maxwell-Boltzmann distributions. The dashed line
represents oscillator $1$ and the solid one represents oscillator $2$. Here, $k=\hbar=t_0=1$,
$(kt_0/\hbar)T_1(0)=8$, $(kt_0/\hbar)T_2(0)=2$, $\omega_1t_0=10$, $\omega_2t_0=4$, $\omega t_0=5$, and
$\tau/t_0=1.5$. Again, the new equilibrium distribution does not satisfy Maxwell-Boltzmann statistics and no
temperature can be assigned to it.} \label{fig:Figure4}
\end{figure}

\newpage

We note some interesting results. The rate with which the equilibrium is reached depends on an interplay between
the characteristic time interval $\tau$ and the interaction frequency $\omega$. Comparing Figs.
\ref{fig:FigureLargeTau} and \ref{fig:FigureSmallTau}, we observe that, for a constant interaction frequency and
a shorter characteristic time interval, the harmonic oscillator system reaches equilibrium much faster. This can
be understood by noting that as $\tau$ decreases, the period of the applied interaction Hamiltonian is
decreased, and the oscillators do not have as much time to interact before the interaction is interrupted.
Decreasing the length of the interaction prevents the oscillators from returning back the temperature they have
exchanged so the equilibrium is achieved faster. The same effect is observed when we compare Figs.
\ref{fig:FigureLargeTau} and \ref{fig:FigureHighFrequency}: for a constant characteristic time interval and a
higher interaction frequency, the oscillators are forced to exchange temperature much faster. Again, the
equilibrium is attained more quickly.

Further, we note that our prediction is verified: $\omega_1\theta_1(\infty)=\omega_2\theta_2(\infty)$ (or,
$\omega_1/[kT_1(\infty)]=\omega_2/[kT_2(\infty)]$), at equilibrium. We observe that when $\omega_1=\omega_2$,
the behavior of the system is symmetric. The two oscillators exchange an equal amount of energy amongst
themselves per unit time. The interaction Hamiltonian takes both oscillators out of their respective equilibria
and, eventually, after many iterations of the process, the two attain an equilibrium Maxwell-Boltzmann
distribution at an effective temperature $T_{1,2}(\infty)=\left[T_1(0)+T_2(0)\right]/2$ that is different from
their initial temperatures. The equilibrium reached here is a thermal equilibrium, in other words, the
oscillators have reached a state where their temperatures have ceased to change, and a single temperature can be
attributed to the entire system. This was in fact noted by several individuals including Montroll and Shuler
\cite{Montroll}, Mathews, Shapiro, and Falkoff \cite{Mathews}, as well as Rau \cite{Rau}, but the availability
of numerical techniques for carrying out the iteration was not present at the time. Nonetheless, the predictions
made were correct and the results are not surprising. They are simply a consequence of the zeroth law of
thermodynamics: when two systems are put in contact with each other, there will be a net exchange of energy
between them unless or until they are in thermal equilibrium. This phenomenon has also been studied by Andersen
and Shuler for the specific case of the relaxation of a hard-sphere Rayleigh and Lorentz gas \cite{Andersen}.

In all of the above examples, the frequencies of the oscillators were equal. A somewhat more unexpected result
(from a thermodynamical point of view), appears in the case where $\omega_1\neq\omega_2$. Here, the exchange of
energy in the system is not symmetric, such behavior being caused by the fact that the frequencies are not
equal. The equilibrium reached is not a thermal one and the final combined distribution no longer satisfies
Maxwell-Boltzmann statistics. The oscillators interact through the interaction Hamiltonian which takes them out
of their initial equilibrium, they exchange quantities (including temperature), and eventually reach a state
where each oscillator has attained a new equilibrium at it's own temperature, such that
$\omega_1/[kT_1(\infty)]=\omega_2/[kT_2(\infty)]$. Besides imposing the condition that the total number of
particles is a constant, the interaction Hamiltonian also dictates that the free energy of the system (the
energy not including the interaction energy) is conserved only when the frequencies of the oscillators are
equal. Here, the selection rules imposed by the interaction Hamiltonian override the statistical mechanical
effects.

\section{Conclusions and Beyond}
We have explicitly calculated the time evolution of an initially uncoupled system of two harmonic oscillators,
under the repeated application of an interaction Hamiltonian for successive time intervals $\tau$. Our results
were calculated exactly using iterative methods and without using perturbation theory. We started with two
initially uncoupled quantum harmonic oscillators, each at equilibrium and with temperatures $T_1$ and $T_2$
respectively, and showed that after enough periodically repeated applications of the interaction Hamiltonian,
the oscillators came to equilibrium with each other - \emph{if} their frequencies were equal - at an effective
temperature which was different from their initial temperatures. When the frequencies were unequal, each
oscillator attained a new equilibrium but there was no effective temperature that could be attributed to both of
them. On the one hand, when the frequencies of the oscillators were the same, the system's initial
Maxwell-Boltzmann distributions evolved into a thermal equilibrium at a new Maxwell-Boltzmann distribution
through a series of transient Boltzmann distributions and on the other hand, when the oscillator frequencies
were unequal, the equilibrium reached was not thermal and the final combined distribution at equilibrium no
longer satisfied Maxwell-Boltzmann statistics. To explain this, we showed that, at equilibrium,
$\omega_1\theta_1(\infty)=\omega_2\theta_2(\infty)$. Further, we have concluded that the selection rules imposed
by the interaction Hamiltonian override statistical mechanical effects.

An important ingredient of our model is what we have called the ``refresh'' procedure. After applying a constant
interaction Hamiltonian to our initially uncoupled set of harmonic oscillators, we perform an act of
``refreshing''. This corresponds to the assumption that there are no fixed phase relationships between the two
harmonic oscillators, while at the same time, the phase relationships between the number states of each
individual oscillator are definite. In addition, we assume that we have no knowledge about the occupation
numbers of each of the modes in the second system and so we average over all these occupation numbers. The fact
that the quantity $N=n_1+n_2$ is conserved by definition of the total Hamiltonian in our model, and the fact
that the initial density matrix is diagonal in the number basis $\{|n_1n_2\rangle\}$ of the harmonic
oscillators, forces the evolved reduced density matrices $\rho_{1}(n\tau)$ and $\rho_{2}(n\tau)$ with
$n=0,1,2,...,\infty$, to be diagonal in their respective number bases $\{|n_1\}$ and $\{|n_2\}$. In turn, this
allows us to write the new combined density matrix after the ``refreshing'' as a product state of the new
reduced states of the two oscillators. We have shown how the ``refresh'' procedure can be obtained through a
single, non-selective measurement on one of the two oscillators. A single spectrum measurement in the number
basis of oscillator $1$, results in the diagonalization of the output state in both oscillator number bases
$\{|n_1\rangle\}$ and $\{|n_2\rangle\}$, and this in effect permits us to assume statistical independence
between the two post-measurement states by allowing us to write the global output state as a tensor product of
$\rho_1$ and $\rho_2$. Even though the transformation of the ``refresh'' procedure formally appears to be
nonlinear in $\rho_{12}$, this is simply a consequence of how we choose to express it. In effect, the
``refreshing'' model in no different than performing a spin measurement on an entangled electron pair.
Naturally, the results obtained in this paper are dependent on the specific model presented, and more
specifically, the interaction Hamiltonian and the ``refresh'' procedure. The question we may now ask is:
\emph{How will our results change if any or both of these factors are changed?}

\section*{ACKNOWLEDGEMENTS}
One of the authors (A.~C.) would like to express her sincere thanks to Anil Shaji, C\'{e}sar A. Rodr\'{i}guez,
Kavan Modi, Todd Tilma, Yafis Barlas, and Andrew Randono for their valuable help on several insightful
discussions and for proofreading the manuscript.

\end{document}